\documentclass[aps,prb,preprint,citeautoscript,superscriptaddress,floatfix]{revtex4}
\usepackage{amsmath,graphicx}
\usepackage{bm}

\newcommand{\mc}{$\mu  \mbox{C} \cdot \mbox{cm}^{-2}$}
\newcommand{\ymno}{YMnO$_3$}

\begin{document}

\title{Landau theory of topological defects in multiferroic hexagonal manganites}
\author{Sergey Artyukhin}
\affiliation{Zernike Institute for Advanced Materials, University of Groningen, Nijenborgh 4, 9747 AG, Groningen, The Netherlands}
\author{Kris~T.~Delaney}
\affiliation{Materials Department, University of California, Santa Barbara, California 93106-5050, USA}
\author{Nicola~A.~Spaldin}
\affiliation{Materials Theory, ETH Zurich, Wolfgang-Pauli-Strasse 27, CH-8093 Zurich, Switzerland}
\author{Maxim Mostovoy}
\affiliation{Zernike Institute for Advanced Materials, University of Groningen, Nijenborgh 4, 9747 AG, Groningen, The Netherlands}

\begin{abstract}

Topological defects in ordered states with spontaneously broken symmetry often have unusual physical properties, such as fractional electric charge or a quantised magnetic field-flux, originating from their non-trivial topology.
Coupled topological defects in systems with several coexisting orders give rise to unconventional functionalities, such as the electric-field control of magnetisation in multiferroics resulting from the coupling between the ferroelectric and ferromagnetic domain walls.
Hexagonal manganites provide an extra degree of freedom: In these materials, both ferroelectricity and magnetism are coupled to an additional, non-ferroelectric structural order parameter. 
Here we present a theoretical study of topological defects in hexagonal manganites based on Landau theory with parameters determined from first-principles 
calculations.
We explain the observed flip of electric polarisation at the boundaries of structural domains, the origin of the observed discrete vortices, and the clamping between ferroelectric and antiferromagnetic domain walls. 
We show that structural vortices induce magnetic ones and that, consistent with a recent experimental report, ferroelectric domain walls can carry a magnetic moment.

\end{abstract}
\maketitle


Recent experimental and theoretical advances in the field of multiferroics have clarified the microscopic 
mechanisms of coupling between ferroelectricity and magnetism in \emph{bulk} materials and led to the discovery of many families of materials in which ferroelectricity is induced by a spin ordering. 
Examples include the orthorhombic rare-earth manganites, spinels, hexaferrites and delafossites, which usually 
have spin orderings of the cycloidal or conical spiral type\cite{Cheong_NatMat_2007,Ramesh_NatMat_2007,Khomskii_Physics_2009}.
The resulting electric polarisation is highly susceptible to an applied magnetic field and can be easily 
rotated or reversed\cite{Kimura_PRL_2003,Hur_Nature_2004,Kitagawa_NatMat_2010}. 
However, this magnetically-induced polarisation is usually too small to allow manipulation of spin states 
by an applied voltage.
Much larger electric polarisations are found in multiferroics such as BiFeO$_3$ and the hexagonal 
rare-earth manganites, in which ferroelectricity results not from spin ordering but from 
electronic and lattice instabilities\cite{Lebeugle_APL_2007,Neaton_PRB_2005,Aken_NatMat_2004}.
Yet in these materials, the electric control of magnetism is not 
straightforward\cite{Chu_NatMat_2008,Lebeugle_PRL_2009,Skumryev_PRL_2011}, since the direction of spins in the magnetically 
ordered state is not correlated with the sign of the macroscopic electric polarisation\cite{Fennie_PRL_2008}.

While enhancing {\it bulk} couplings between polarisation and magnetism is difficult, 
practical switching of a ferroic order parameter with an applied field invariably involves 
motion of the domain walls.
Magnetoelectric {\it switching} therefore depends crucially on interactions between
ferroelectric and ferromagnetic domain walls, which are not as well understood.
In this context, the observed clamping between ferroelectric 
and antiferromagnetic domain walls in multiferroic hexagonal manganites\cite{Fiebig_Nature_2002} provides
a unique prototype for investigation.

\begin{figure}[htbp]
 \centering
\includegraphics[width=0.77\columnwidth]{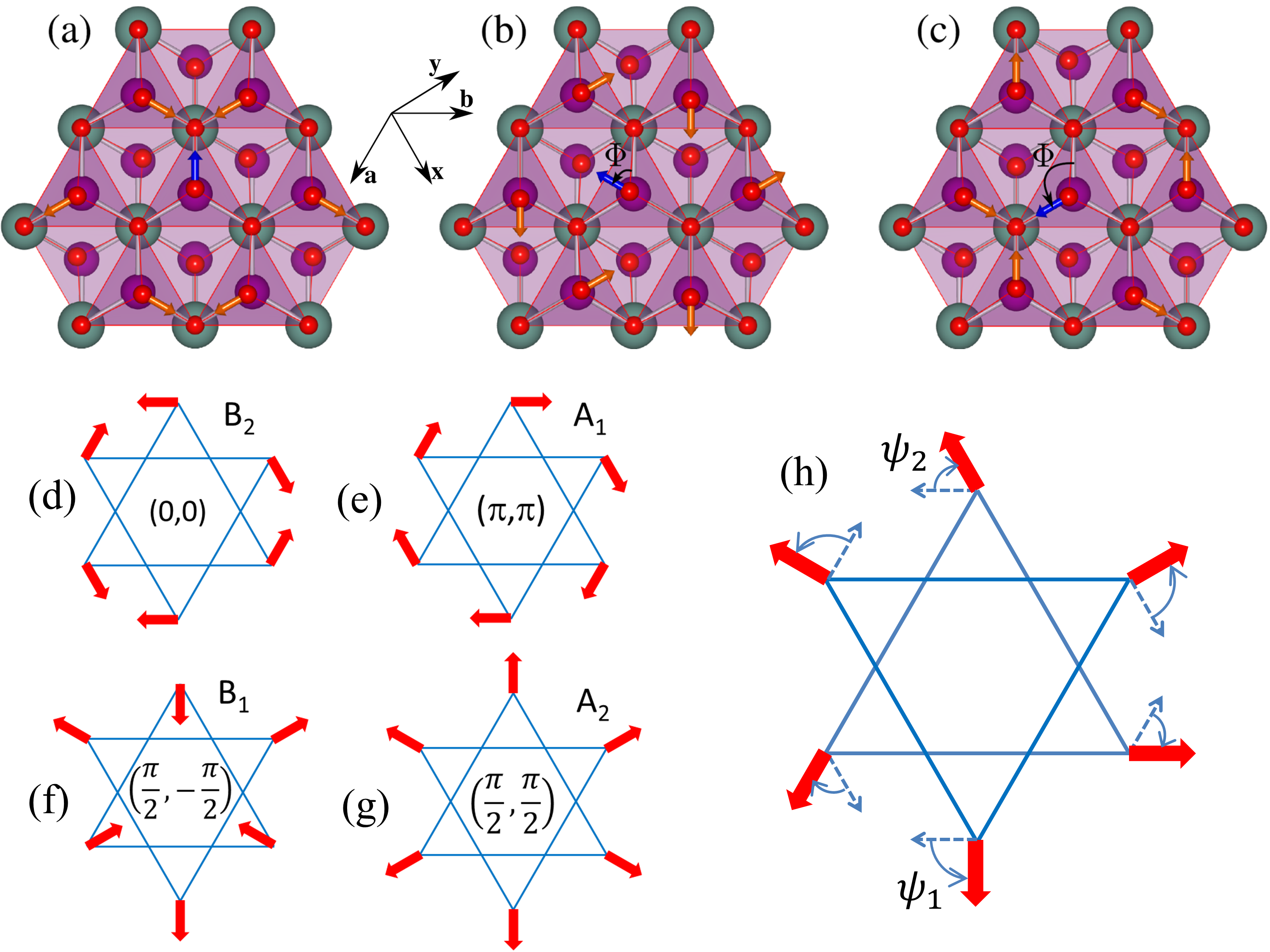}
\caption{{\bf Structural and magnetic angles.} {\bf a},{\bf b},{\bf c}, Projections of the displacements of apical oxygen ions on the $ab$ plane in the trimerised state indicated by arrows. One ion (blue arrow) is chosen to define the trimerisation phase $\Phi$. Shown are the $\alpha_+$ state with $\Phi=0$ ({\bf a}), $\gamma_{-}$ state with $\Phi = \pi/3$ ({\bf b}) and the $\beta_{+}$ state with $\Phi=\frac{2\pi}{3}$ ({\bf c}). Also shown are the displacements of apical oxygen ions in an adjacent Mn-O layer (lighter triangles). {\bf d},{\bf e},{\bf f},{\bf g}, The four magnetic states of hexagonal manganites with the spin directions indicated by red arrows and the corresponding values of the angles $(\chi_1,\chi_2)$, where $\chi_{1,2} = \psi_{1,2} - \Phi$ (see text). {\bf h}, The angles $\psi_1$ and $\psi_2$ describing the rotations of spins in magnetic domain walls ($\psi_1,\psi_2>0$ correspond to the clockwise(anticlockwise) rotation in even(odd) Mn layers).}
 \label{fig:trim}
\end{figure}

The hexagonal manganites, $R$MnO$_3$, where $R$ denotes a small-radius rare earth ion (Dy, Ho, Er, Tm, Yb, Lu), Y or Sc, are improper ferroelectrics;
electric polarisation appears as a by-product of a primary structural transition\cite{Aken_NatMat_2004}. 
The crystal structure consists of corner-sharing MnO$_5$ trigonal bipyramids, which form triangular layers, 
separated by layers of $R$ ions. 
The structural transition above $1000$\,K results in periodic tilts of the MnO$_5$ bipyramids and displacements of the 
$R$ ions along the $c$ axis normal to the layers\cite{Yakel_AC_1963,Katsufuji_PRB_2001}.
This periodic $\sqrt{3} \times \sqrt{3}$ lattice distortion makes the size of the unit cell three 
times larger and is often referred to as the trimerisation transition (see Fig.~\ref{fig:trim}(a-c)).
The anharmonic coupling between the trimerisation mode and a polar optical phonon mode induces the observed 
electric polarisation along the $c$ axis, $P_c \sim 5.5$\,\mc\cite{Fennie_PRB_2005}.

At much lower temperatures, $\sim100$\,K, an antiferromagnetic ordering of Mn spins emerges. 
While there is a large body of evidence for the strong interplay between the microscopic spin, charge and lattice 
degrees of freedom in hexagonal manganites\cite{Katsufuji_PRB_2001,Cruz_PRB_2005,Lee_Nature_2008,Adem_JPCM_2009}, 
the sign of the overall antiferromagnetic order parameter is decoupled from the direction of the macroscopic 
electric polarisation, as such correlation is forbidden by symmetry. 
It therefore came as a surprise when non-linear optical measurements of \ymno\ demonstrated that ferroelectric 
domain walls are locked to magnetic ones\cite{Fiebig_Nature_2002}.
Furthermore, this clamping was found to be non-reciprocal, as ``free'' magnetic domain walls, not associated 
with the electric polarisation reversals, were also observed. 
Proposals for the clamping mechanism have included strain mediation\cite{Goltsov_PRL_2003} and renormalisation of spin interactions at ferroelectric walls\cite{Hanamura_JPCM_2003,Hanamura_JPSJ_2003}.

Recently, new observational evidence has shed light on the mechanism for clamping.
A combination of conducting atomic-force microscopy and transmission-electron microscopy demonstrated that the ferroelectric 
domain walls are pinned to the boundaries of the structural domains that appear upon transition to the trimerised state.\cite{Choi_NatMat_2010,Chae_PNAS_2010} 
These measurements also revealed intricate patterns of unusual line defects. 
These so-called `cloverleaf' defects,\cite{Choi_NatMat_2010,Chae_PNAS_2010}  at which six different structural and ferroelectric domains merge, have also been seen in piezoresponse 
force microscopy.\cite{Jungk_APL_2010,Lilienbaum_APL_2011} 
In Ref.~\onlinecite{Mostovoy_NatMat_2010} it was suggested that the line defects are discrete 
analogues of vortices, and that the change of polarisation sign at structural domain boundaries is
a consequence of the special form of the coupling between the lattice distortion and electric polarisation originating from the ``geometric'' nature of ferroelectricity in 
hexagonal manganites\cite{Aken_NatMat_2004,Fennie_PRB_2005}.
These results put the clamping between the ferroelectric and antiferromagnetic 
domain walls into an entirely new perspective.   

In this paper we study the interplay between structural, ferroelectric and magnetic defects in hexagonal manganites using an expansion of the free energy in powers of the corresponding order parameters and their gradients.
Based on symmetries of ordered states of these materials, this approach allows us to identify stable topological defects and describe their mutual interactions in the most economical way.
In Ref.~\onlinecite{Fennie_PRB_2005} Fennie and Rabe discussed the Landau theory of improper ferroelectricity for spatially uniform states of hexagonal manganites.
Extracting parameters of the Landau expansion from first-principles studies of YMnO$_3$, they showed that polarisation
emerges due to a non-linear coupling to the trimerisation mode.
We extend this theory to inhomogeneous topological defects and include spin degrees of freedom to study effects of the structural domain walls and vortices on magnetic ordering.

In the trimerised state three neighbouring MnO$_5$ bipyramids tilt towards (or away from) their common equatorial oxygen 
atom (see Fig.~\ref{fig:trim}(a-c))\cite{Aken_NatMat_2004}.
As a consequence of the hexagonal structure of Mn-O layers, there are six distinct trimerised states, corresponding to six degenerate 
minima of the lattice energy. 
Being a periodic lattice modulation in a layered system, the trimerisation is described entirely by the amplitude $Q$ and phase $\Phi$. 
The physical meaning of the phase $\Phi$ is the azimuthal angle describing the in-plane displacements of apical 
oxygens (see Fig.~\ref{fig:trim}(a-c)). 
The minimal-energy states can then be labelled by the six values of the phase: $0,\pm\frac{\pi}{3},\pm\frac{2\pi}{3}$ and $\pi$. 
At the structural domain boundaries $\Phi$ varies spatially between two of these six values.

Microscopically, the trimerisation is the condensation of the zone-boundary $K_3$ mode with wave vector $\mathbf{q} = (1/3,1/3,0)$, which breaks the P$6_3/mmc$ symmetry of the undistorted phase lowering it to P$6_3cm$.  
Similarly, the spontaneous electric polarisation $P_c$ is proportional to the amplitude of the zone-centre mode ${\cal P}$ with $\Gamma_2^{-}$ symmetry.  
This polar mode is stable in the P$6_3/mmc$ structure, but is non-linearly coupled to the unstable $K_3$ mode, and therefore appears together with the trimerisation.

\begin{table}[htbp]
\centering
\begin{tabular}{|c|c|c|c|c|c|c|}
\hline
&$S_{a}$&$3_c$&$\tilde{2}_c$&$m_{a+b}$&$I$&$T$\\
\hline
$\Phi$&$\Phi+2\pi/3$&$\Phi$&$-\Phi$&$-\Phi$&$\pi-\Phi$&$\Phi$\\
$\psi_1$&$\psi_1+2\pi/3$&$\psi_1$&$\pi-\psi_2$&$-\psi_1$&$-\psi_2$&$\psi_1+\pi$\\
$\psi_2$&$\psi_2+2\pi/3$&$\psi_2$&$\pi-\psi_1$&$-\psi_2$&$-\psi_1$&$\psi_2+\pi$\\
$P_c$&$+P_c$&$+P_c$&$+P_c$&$+P_c$&$-P_c$&$+P_c$\\
$H_c$&$+H_c$&$+H_c$&$+H_c$&$-H_c$&$+H_c$&$-H_c$\\
[0.4ex]
\hline
\end{tabular}
\caption{Transformations of the trimerisation phase $\Phi$, the spin angles $\psi_1$ and $\psi_2$, the electric polarization $P_c$ and the magnetic field $H_c$ under the generators of the P$6_3/mmc$ space group describing the high-temperature phase: translation $S_a = (x+1,y,z)$, three-fold axis  $3_c = (-y,x-y,z)$, two-fold screw axis ${\tilde 2}_c = (-x,-y,z+1/2)$, mirror plane $m_{a+b} = (-y,-x,z)$, inversion $I = (-x,-y,-z)$, and the time reversal operation $T$.}
\label{tab:transformation}
\end{table}

The free-energy expansion in powers of $Q,{\cal P}$, and their gradients,
\begin{eqnarray}
f = &~& \frac{a}{2}Q^2 + \frac{b}{4} Q^4 + \frac{Q^6}{6}\left(c+c'\cos 6\Phi\right)\nonumber\\ 
&-& g Q^3 {\cal P} \cos 3\Phi +\frac{g'}{2}Q^2 {\cal P}^2 + \frac{a_{{\cal P}}}{2}{\cal P}^2\\
&+&\frac{1}{2} \sum_{i = x,y,z}\left[s_Q^i \left(\partial_i Q\partial_i Q + Q^2 \partial_i \Phi \partial_i \Phi\right) +
s_{\cal P}^i \partial_i {\cal P}\partial_i {\cal P}\right],\nonumber 
\label{eq:Fu}
\end{eqnarray}
is obtained using the transformation  properties of the trimerisation phase $\Phi$ and the polarisation $P_c$ under the generators of the high-temperature space group summarised in Table \ref{tab:transformation}. We consider only the lowest-order stiffness terms accounting for the energy cost of spatial variations of $Q$ and ${\cal P}$. $(x,y)$ are the Cartesian coordinates in the $ab$ plane (see Fig.~\ref{fig:trim}), and by symmetry, $s_Q^x = s_Q^y$ and $s_{\cal P}^x = s_{\cal P}^y$.

The trimerisation phase $\Phi$ and the stiffness terms, $s_Q$ and $s_{\cal P}$, not considered by Fennie and Rabe\cite{Fennie_PRB_2005}, play an important role in the theory of topological defects.  
In particular, the form of the non-linear coupling, $-g Q^3 {\cal P} \cos 3\Phi$, giving rise to improper ferroelectricity, implies that for $g>0$ the electric polarisation induced in the 
states with $\Phi = 0, +\frac{2\pi}{3}$ and $-\frac{2\pi}{3}$ is positive 
(the $\alpha_+$, $\beta_+$ and $\gamma_+$ phases\cite{Choi_NatMat_2010,Mostovoy_NatMat_2010}), 
while for $+\frac{\pi}{3},\pi$ and $-\frac{\pi}{3}$ it is negative (the $\gamma_-$, $\alpha_-$ and $\beta_-$ phases). 
In other words, neighbouring trimerisation phases, separated by $\Delta \Phi = \frac{\pi}{3}$, have opposite electric polarisations.

\begin{table}[htbp]
\centering
\begin{equation*}
\begin{array}{cclccclccclcccl}
a &=& -2.626 \;\mbox{eV} \cdot \mbox{\AA}^{-2}&\;\;&b &=& 3.375 \;\mbox{eV} \cdot \mbox{\AA}^{-4}&\;\;&
c &=& 0.117 \;\mbox{eV} \cdot \mbox{\AA}^{-6}&\;\;&c' &=& 0.108 \;\mbox{eV}\cdot \mbox{\AA}^{-6}\\
a_P &=& 0.866 \;\mbox{eV} \cdot \mbox{\AA}^{-2}&\;\;&g &=& 1.945\; \mbox{eV}\cdot \mbox{\AA}^{-4}&\;\;&
g' &=&  9.931\; \mbox{eV}\cdot \mbox{\AA}^{-4}&\;\;& & &\\
s^{z}_{Q} &=& 15.40 \;\mbox{eV} &\;\;&s^{x}_{Q} &=& 5.14\; \mbox{eV}&\;\;&
s^{z}_{P} &=& 52.70 \;\mbox{eV}&\;\;\;&s^{x}_{P} &=& -8.88\; \mbox{eV}
\end{array}
\end{equation*}
\caption{Parameters of the phenomenological expansion of the free energy Eq. (\ref{eq:Fu})  obtained from \textit{ab initio} calculations. All parameters are calculated per unit cell of the trimerised lattice with the volume $V = 365.14$\,\AA$^{3}$ containing 6 formula units. The polarization $P_c$ is related to the amplitude of the polar mode ${\cal P}$ by 
$P_c = V^{-1}\bar{Z}^\star\mathcal{P}$, where $\bar{Z}^\star = 9.031 e$ is the effective charge of the polar mode. Because of the negative stiffness, $s^{x}_{P}$, of the polar mode, the term $\frac{1}{2} t^{x}_{P} \left[\left(\frac{\partial^2}{\partial x^2} + \frac{\partial^2}{\partial y^2}\right){\cal P}\right]^2$ with $t^x_P = 73.56 \mbox{eV} \cdot \mbox{\AA}^{2}$ was added to Eq.(1) in order to calculate the structure of the domain wall shown in Fig.~\ref{fig:PhiDW}(a). }
\label{tab:parameters}
\end{table}


The numerical values of the parameters $a$, $b$, etc. in Eq. (\ref{eq:Fu}) for YMnO$_3$ are listed in Table~\ref{tab:parameters}.
In order to determine them, we performed \emph{ab initio} supercell calculations for the various lattice distortions (see Appendix A).
In particular, Fig.~\ref{fig:dft_homogfits}(a) shows the variation of the total energy with respect to the magnitude of $Q$ for 
two chosen values of $\Phi$ and ${\cal P} = 0$. 
It is clear that in the physically relevant range of $Q$, the total energy depends very weakly on $\Phi$. 
The $\Phi$-dependence of the energy $f(Q,\Phi,{\cal P}=0)$ first appears in the sixth-order of the expansion in powers of $Q$ and the corresponding coefficient $c^\prime$ is small.
The energy landscape in the $(Q,\Phi)$ plane therefore essentially has a rotationally invariant Mexican Hat shape with no barriers separating the six structural domains.

\begin{figure}[htbp]
\centering
\includegraphics[width=0.8\columnwidth]{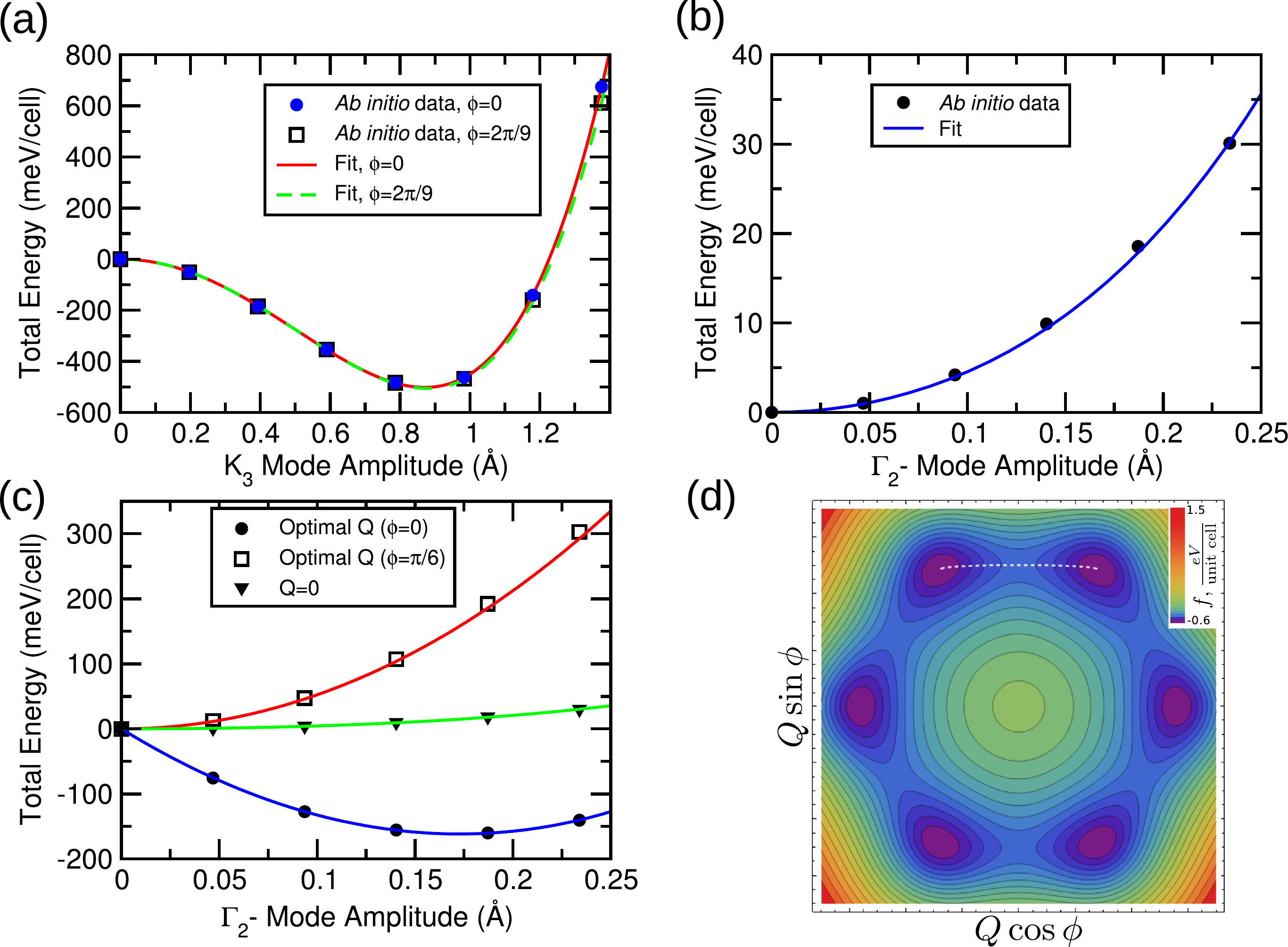}
\caption{{\bf Extraction of model parameters from \emph{ab initio} calculations.}   
{\bf a}, Variation of total energy with respect to $K_3$ mode amplitude for two trimerisation angles, $\Phi=0$ and $\Phi=\frac{2\pi}{9}$. 
Parameters $a$, $b$, $c$ and $c^\prime$ are extracted, and owing to the weak angular dependence of the energy $c^\prime$ is small.
{\bf b}, Variation of total energy about the high-symmetry P6$_3$/mmc structure
with respect to the amplitude of the $\Gamma_2^-$ (polar) mode, $\mathcal{P}$. The polar mode is stable and does not spontaneously emerge at $T=0$\,K.
{\bf c}, Coupling between $K_3$ and $\Gamma_{2-}$ modes for two different trimerisation angles. 
For trimerisation angles of $0, \pm\frac{\pi}{3}, \pm\frac{2\pi}{3}$ and $\pi$ the anharmonic 
coupling leads to a non-zero polarisation and a total energy lowering of $\sim 26$\,meV per formula unit. On the other hand, intermediate trimerisation angles do not allow polarisation to develop.  The $g$ and $g^\prime$ parameters are extracted from these data.
{\bf d}, Contour plot of the free energy of uniformly trimerised states as a function of $Q$ and $\Phi$. Here, $P_z$ has been optimized for each $Q$, $\Phi$. The trajectory $Q(\Phi)$ (white dashed line) connecting two neighbouring energy minima corresponds to the lowest-energy structural domain wall.}
\label{fig:dft_homogfits}
\end{figure}

In reality, the $\Phi$-rotation of the tilted bipyramids is not a zero mode, because an additional $\cos 6 \Phi$ term is generated by minimising $f$ with respect to ${\cal P}$ and eliminating ${\cal P}$ from Eq.(\ref{eq:Fu}), which lowers the energy by 
$
\frac{g^2 Q^6 (\cos 6\Phi + 1)}{4(g'Q^2 + a_{{\cal P}})}.
$
Fig.~\ref{fig:dft_homogfits}(c) shows the dependence of the total energy on ${\cal P}$ for $\Phi = 0$ (at which the energy has minimum) and $\Phi = \pi/6$ (at the top of the barrier separating two minimal-energy states). 
Clearly, the anharmonic coupling between $Q$ and ${\cal P}$ results in a strong $\Phi$ dependence of the energy. 
Therefore, even though the emergence of ferroelectricity in hexagonal manganites is improper, the coupling of trimerisation to polarisation 
is the only factor that determines the energetic barriers between different trimerised states, and is responsible for replacing the 
accidental continuous $XY$ symmetry of $f$ with the discrete Z$_6$ symmetry.

\begin{figure}[htbp]
\centering
 \includegraphics[width=\columnwidth]{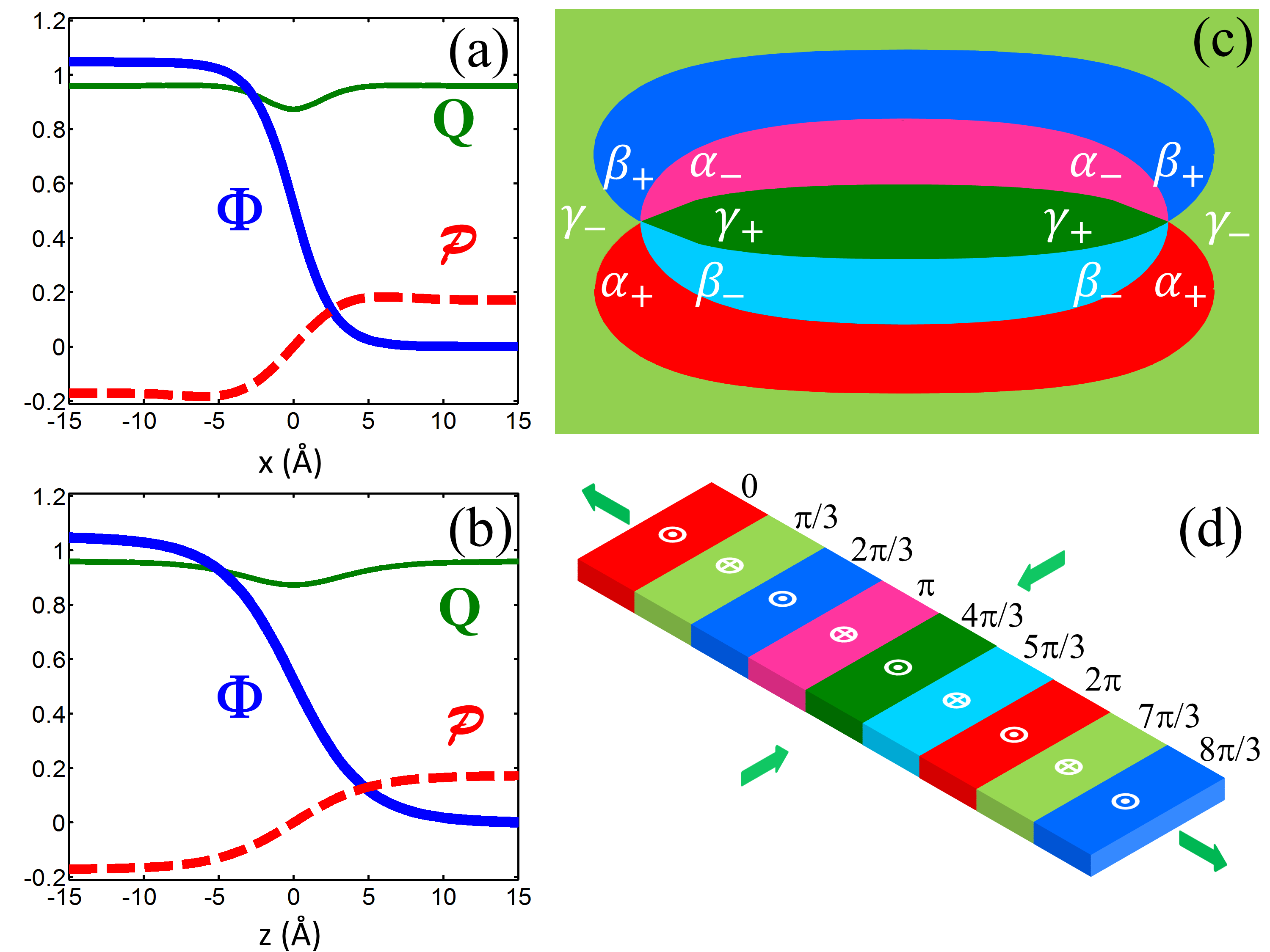}
 \caption{{\bf Structural topological defects.} {\bf a},{\bf b}, The variation of the trimerisation angle $\Phi$ (measured in radians, thick blue line), trimerisation amplitude $Q$ (measured in \AA, thin green line) and the polar mode ${\cal P}$ (measured in \AA, dashed red line) across the lowest-energy domain wall normal to the $ab$ plane ({\bf a}) and parallel to the $ab$ plane ({\bf b}). 
{\bf c}, The vortex-antivortex pair. In the structural vortex/antivortex the angle $\Phi$  increases/decreases by $2\pi$ along the loop encircling the vortex core in the counter-clockwise direction (the 6 trimerisation states are indicated by colour). {\bf d}, The `topological stripe domain state' in a thin film of ferroelectric hexagonal manganite parallel to the $ab$ plane with the alternating polarisation (white symbols) along the $c$ axis. A strain indicated by green arrows results in the monotonic increase of the trimerisation angle in the direction normal to the stripes. }
\label{fig:PhiDW}
\end{figure}

This has a strong effect on the structure of topological defects in the trimerised state, which can be described as trajectories in the $(Q,\Phi)$-plane minimising the energy for given initial and final conditions. 
For example, a structural domain wall corresponds to a path connecting two energy minima. 
The shortest path connecting two `neighbouring' states whose trimerisation angles differ by  $\Delta \Phi = \pm\pi/3$ (see Fig.~\ref{fig:dft_homogfits}(d)) is the lowest-energy domain wall. This path follows the bottom fold of the Mexican Hat where the potential barrier between the two minima is the lowest, so that the amplitude of the trimerisation $Q$ in the wall is close to its bulk value. Figures~\ref{fig:PhiDW}(a-b) show the coordinate dependence of $\Phi$ and $Q$ across domain walls obtained by numerical free-energy minimisation. The domain wall width is $5-10$ \AA\ and $Q$  at the domain wall is reduced by about 10\%.

Since the neighbouring energy minima separated by $\Delta \Phi = \pm \pi / 3$ have anti-parallel electric polarisations, the structural domain wall is at the same time a ferroelectric domain wall (see Fig.~\ref{fig:PhiDW}(a-b)). The improper nature of ferroelectricity in hexagonal manganites forbids purely ferroelectric domain walls, i.e. the reversals of $P_c$ within one structural domain, since  the sign of $P_c$ is uniquely determined by the sign of $\cos 3\Phi$. Furthermore, the structural domain walls with $\Delta \Phi  = 2\pi/3$, separating states with the same electric polarisation, are unstable: they `decay' into lowest-energy domain walls with $\Delta \Phi  = \pi/3$ which we find to repel each other when they overlap. This explains the clamping between the structural and ferroelectric domain walls observed by Choi {\em et al.}\cite{Choi_NatMat_2010}

Spherical or cylindrical structural domains are unstable against shrinking, which explains the scarcity of closed domain wall lines in hexagonal manganites.\cite{Chae_PNAS_2010,Jungk_APL_2010} To be stable, the domain walls must terminate either at the surface of the sample or at another type of stable topological defect -- the structural vortex or anti-vortex, shown in Fig.~\ref{fig:PhiDW}(c). At the vortex line,  where the trimerisation amplitude $Q$ vanishes,  all six structural domains meet in such an order that the trimerisation phase $\Phi$ changes by $2 \pi$ around a contour encircling the vortex line.\cite{Mostovoy_NatMat_2010} Such a defect cannot be unwound and can only be annihilated by an anti-vortex, around which the phase changes by $- 2\pi$ (see Fig.~\ref{fig:PhiDW}(c)).  Away from the core, the trimerisation phase $\Phi$ varies strongly only at the six radial domain walls. The electric polarisation changes sign at each domain wall and varies six times along a loop encircling the vortex line. These vortices and anti-vortices are the `cloverleaf defects' observed in Ref.~\onlinecite{Choi_NatMat_2010}.

Fig.~\ref{fig:PhiDW}(c) shows a vortex--anti-vortex pair configuration obtained by minimising the energy for a given distance between these defects. The domain walls diverge radially from the vortex/anti-vortex core with the $60^\circ$-angle between neighbouring domain walls. Far from the core they bend and become parallel to minimise the total length of the structural boundaries, which gives rise to a linear potential between the discrete vortices as opposed to the  logarithmically growing potential for continuous vortices.\cite{ChaikinLubensky} Despite this  confining potential, vortex lines and domain walls form dense networks\cite{Choi_NatMat_2010,Chae_PNAS_2010,Jungk_APL_2010,Lilienbaum_APL_2011} that are snapshots of states close to critical temperature capturing the formation of these topological defects by large thermal fluctuations.\cite{Kaski_PRB_1985,Grest_PRB_1988}

A different type of topologically stable domain pattern is uncovered by considering the lowest-order coupling of the inhomogeneous trimerisation to strains, which has the form
\begin{equation}
f_{\rm strain} = - G Q^2 \left[ \left(u_{xx} - u_{yy}\right)\partial_x \Phi - 2 u_{xy} \partial_y \Phi \right],
\label{eq:fstrain}
\end{equation} 
where $(x,y)$ are the Cartesian coordinates in the $ab$ plane. With such a coupling, a parallel array of structural domain walls, each with the same increment of the trimerisation angle ($\Delta \Phi = \pm \pi/3$), is topologically stable (see Fig.~\ref{fig:PhiDW}(d)). Because of the alternating electric polarisation at the structural walls, such a ``$\Phi$-staircase'' is at the same time a ferroelectric stripe domain state, as stabilised in thin films by the long range dipole-dipole interactions.\cite{Fong_Science_2004,Matsumoto_APL_2011} These interactions are, however, insensitive to the sign of $\Delta \Phi$ at the domain walls, whereas the applied strain selects the direction normal to the walls, in which $\Phi$ increases monotonically.

Next we address the coupling between the magnetism and the structural/ferroelectric domain walls.
The spins on the Mn ions in hexagonal manganites order in one of four different magnetic states: A$_{1}$, A$_2$, B$_1$, and B$_{2}$, shown in Fig.~\ref{fig:trim}(d-g).\cite{Fiebig_APL_2003} Their origin can be understood by considering the hierarchy of interactions
between the magnetic moments on Mn sites. By far the strongest is the antiferromagnetic exchange between neighbouring spins in the triangular layers of Mn ions, which leads to the non-collinear 120$^\circ$ spin ordering. The magnetic easy-plane anisotropy, and the anti-symmetric Dzyaloshinkii-Moriya interaction with Dzyaloshinskii vector along the $c$ axis, confine spins to the $ab$ plane. The local in-plane anisotropy axes, also favouring the $120^{\circ}$ spin angle on neighbouring Mn sites, selects either the ``radial'' (as in the A$_2$ and B$_1$ phases) or the ``tangential'' (as in A$_1$ and B$_2$ phases) orientation of spins (Fig.~\ref{fig:trim}). Finally, the interlayer exchange interactions, which are more than two orders of magnitude weaker than the intra-layer interactions, lead to either even (``A'' phases) or odd (``B'' phases) symmetry under the two-fold screw rotation $\tilde{2}_c$.\cite{Sato_PRB_2003}

\begin{figure}[htbp]
 \includegraphics[width=\columnwidth]{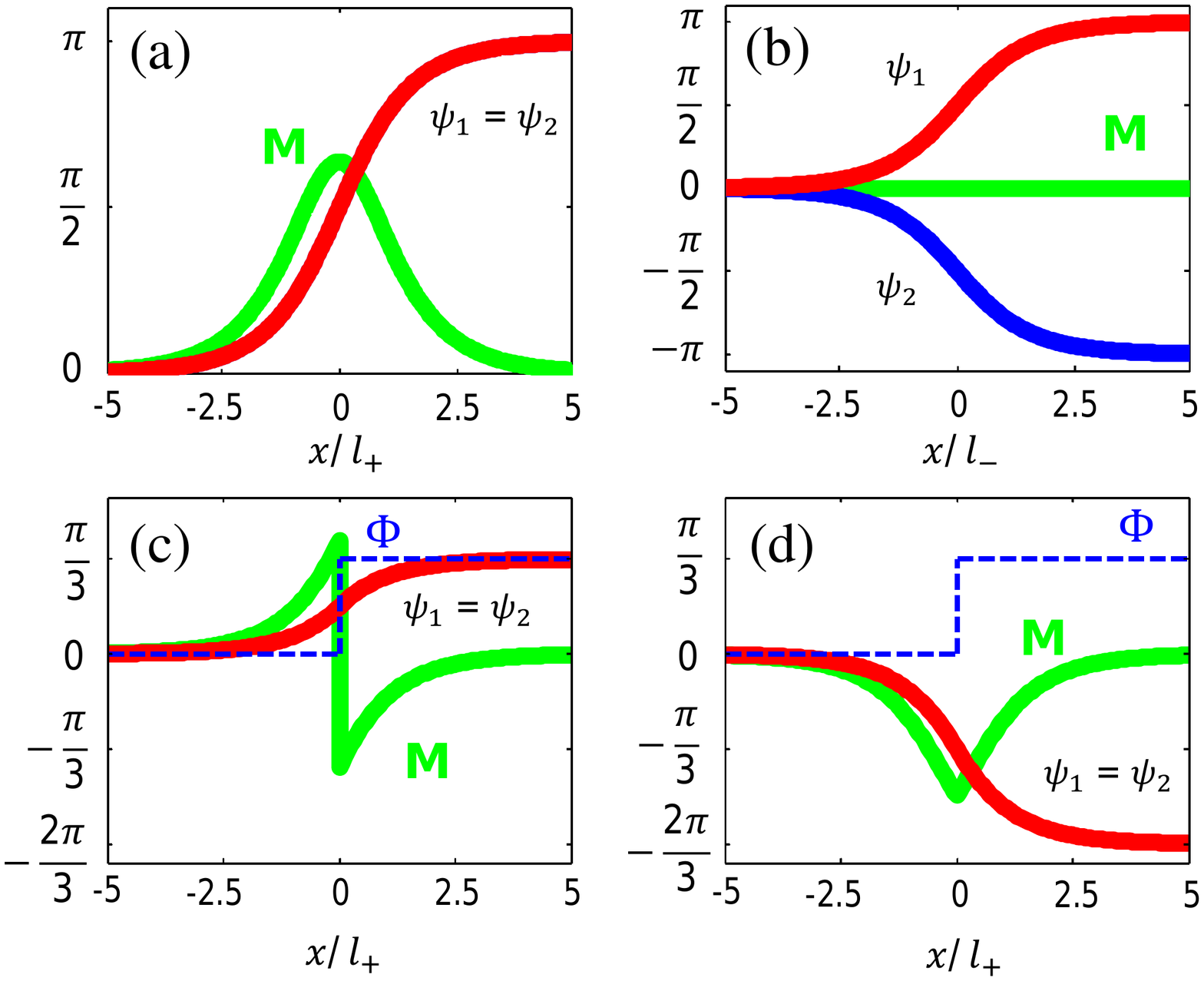}
 \caption{{\bf  Antiferromagnetic domain walls in hexagonal manganites.} {\bf a},{\bf b}, ``Free" $180^{\circ}$ walls within one structural domain. {\bf c},{\bf d}, Domain walls clamped to structural boundaries. Shown is the variation of the spin rotation angles $\psi_1$ and $\psi_2$ (solid red and blue lines, respectively), magnetisation $M_c$ (solid green line, arbitrary units) and the trimerisation phase $\Phi$ (dashed blue line) across the domain wall. At the structural boundary $\Phi$ and the directions of magnetic easy axes on Mn sites change by $60^{\circ}$. The directions of spins adjust over much longer distances by rotating over the $60^{\circ}$ ({\bf c}) or $-120^{\circ}$ ({\bf d}) angle.}
\label{fig:MDW}
\end{figure}

Due to the strong in-plane exchange, the angle between neighbouring spins remains close to 120$^\circ$ throughout a magnetic domain wall. Furthermore, to match the directions of spins on both sides of the wall with the local in-plane anisotropy axes, the spins in the domain wall must rotate in the $ab$ plane. Therefore, the structure of magnetic domain walls can be described by two angles, $(\psi_1,\psi_2)$, quantifying the rotation of spins around the $c$ axis in the even and odd layers respectively,\cite{Fiebig_APL_2003,Goltsov_PRL_2003} as shown in Fig.~\ref{fig:trim}(h).
The transformation properties of $\psi_1$ and $\psi_2$ (see Table \ref{tab:transformation}) determine the form of the magnetic free-energy density: 
\begin{eqnarray}
f_{\rm mag}(\psi_1,\psi_2,H_c) = \!\!\!&~&\!\!\!\!\!\! S\left[ (\partial_\mu \psi_1)^2 +  (\partial_\mu \psi_2)^2 \right] 
+ A\left[\sin^2\chi_1 + \sin^2\chi_2\right]  \nonumber \\ 
&-&\!\!\! C_{+} \cos(\chi_1+\chi_2) - C_{-}\cos(\chi_1-\chi_2) - \frac{1}{2}M_{A_{2}}H_c\left(\sin\chi_1+\sin\chi_2\right),
\label{eq:fmag}
\end{eqnarray}
where $\chi_{1,2} = \psi_{1,2} - \Phi$. The first term originates from the nearest-neighbour exchange. Comparing the energies of the 4 uniform phases (see Fig.~\ref{fig:trim}): $f_{B_{2}} = - C_{+} - C_{-}$,  $f_{A_{1}} = C_{+} + C_{-}$, $f_{A_{2}} = 2A - C_{+} + C_{-}$ and $f_{B_{1}} = 2A + C_{+} - C_{-}$, we conclude that the $A$ coefficient results from the in-plane magnetic anisotropy, $C_{+}$ describes the interlayer exchange interactions, while $C_{-}$ is related to a symmetric exchange anisotropy. The last term in Eq.(\ref{eq:fmag}) describes the weak ferromagnetic moment along the $c$ axis in the A$_2$ phase with $(\chi_1,\chi_2) = \pm (\pi/2,\pi/2)$. Terms proportional to spin operators of power higher than 2 are neglected. Equation~(\ref{eq:fmag}) is used to calculate magnetic structure of the topological defects.

We first consider magnetic domain walls within one structural domain $\Phi = 0$, in which case $\chi_{1,2} = \psi_{1,2}$. The walls separate two magnetic states related by the time reversal operation, so that $\Delta\psi_{1,2} = \pm \pi$ across the wall. Figures~\ref{fig:MDW} (a-b) show two topologically distinct types of such walls between the degenerate B$_2$ states: the one in which spins in neighbouring layers rotate in opposite ($\Delta \psi_1 = \Delta \psi_2 = \pi$) and the same ($\Delta \psi_1 = - \Delta \psi_2 = \pi$) directions. 
The free energies per unit area of the two walls are $8\sqrt{S(A+C_{+})}$ and $8\sqrt{S(A+C_{-})}$ respectively. 
The domain wall with $\psi_1 = \psi_2$ has a nonzero magnetic moment, since the magnetic configuration in the middle of the wall,  $(\psi_1(0),\psi_2(0)) = (\pi/2,\pi/2)$, is of the weakly ferromagnetic A$_2$ type. The net magnetic moment per unit area of the wall is $\pi M_{A_{2}}  l_{+}$, where $M_{A_{2}}$ is the magnetisation in the A$_2$ phase and   $l_{\pm} =  \sqrt{\frac{S}{(A+C_{\pm})}}$ is the domain-wall thickness. 

Importantly, within a domain, the reference triangles of Mn spins that are used to define the four magnetic phases have their apical oxygen ions tilted either towards or away from a common centre (Fig.~\ref{fig:trim}). This is because
the tilts of the oxygen bipyramids, described by the angle $\Phi$, determine the in-plane magnetic anisotropy axes.  A shift by one lattice constant (of the non-trimerised lattice) within the uniform domain results in a $120^{\circ}$ rotation of the anisotropy axes, since $\Phi \rightarrow \Phi + 2\pi/3$. 
To minimise the magnetic energy, therefore, $\psi_1$ and $\psi_2$ must transform in the same way: $\psi_{1,2} \rightarrow \psi_{1,2} + 2\pi /3$. This is why the free energy Eq.(\ref{eq:fmag}) depends on $\chi_{1,2} = \psi_{1,2} - \Phi$, and why, in general, the ``covariant'' angles $(\chi_1,\chi_2)$, rather than $(\psi_1,\psi_2)$, should be used to describe the magnetic phases. 

Now considering structural inhomogeneity, it is evident that the magnetic structure must respond to the presence of a structural domain wall. 
At a structural domain wall with $\Delta \Phi = \pi/3$, for example, spins rotate either by $\pi/3$, in which case $\chi$ is the same on both sides of the wall, or by $-2 \pi/3$, in which case $\Delta \chi = -\pi$. 
Thus, structural domain walls are also magnetic domain walls. 
In the lowest-energy configuration, $\Delta \psi_1 = \Delta \psi_2 = \pi/3$   (see Fig.~\ref{fig:MDW}(c)), while the next-lowest-energy configuration has $\Delta \psi_1 = \Delta \psi_2 = -2\pi/3$ (see Fig.~\ref{fig:MDW}(d)). 
Domain walls in which spins in neighbouring layers rotate over different angles, e.g. $\Delta \psi_1 = \pi/3$ and $ \Delta \psi_2 = - 5\pi/3$, are higher higher in energy.
It is important to stress that the thickness of the magnetic domain walls --- $l_{\pm} \sim 10^2$ \AA\ --- greatly exceeds that of the structural domain wall. The $\Delta \psi = - 2\pi /3$ antiferromagnetic domain wall  clamped to the structural boundary (Fig.~\ref{fig:MDW}(d)) has a nonzero moment along the $c$ axis  equal $-2 \pi/3 M_{A_{2}} l_{+}$  per unit area of the wall. 

With these considerations, we can understand the simultaneous presence of ``clamped'' and ``free'' antiferromagnetic domain walls in hexagonal manganites.\cite{Fiebig_Nature_2002} Every structural domain wall ($\Delta \Phi = \pi/3$) induces a magnetic domain wall in which spins rotate by $60^{\circ}$ or $120^{\circ}$ (Figs.~\ref{fig:MDW} (c-d)). Because of the sign change of electric polarisation at the structural boundary, these antiferromagnetic domain walls appear to be clamped with the ferroelectric domain walls. 
The ``free''  antiferromagnetic domain walls that do not follow ferroelectric domain boundaries are the $180^{\circ}$ antiferromagnetic domain walls  within one structural domain (Figs.~\ref{fig:MDW} (a-b)). 
These results imply that in the minimal-energy magnetic state of the structural vortex spins wind around the vortex core, and the total spin rotation angle along a loop encircling the vortex is $\Delta \psi_1  = \Delta \psi_2 = \Delta\Phi = \pm 2\pi$. That is, structural vortices are also magnetic vortices. 
The small-angle neutron scattering experiment on HoMnO$_3$, the electric switching of magnetisation of coupled ferromagnetic/LuMnO$_3$ thin films, and the magnetic force microscopy study of ErMnO$_3$, all indicated the presence of an uncompensated ferromagnetic moment at antiferromagnetic domain walls.\cite{Ueland_PRL_2010,Skumryev_PRL_2011,Geng_preprint_2012}  Our analysis summarised in Fig.~\ref{fig:MDW} shows that both ``clamped'' and ``free'' antiferromagnetic domain walls  induce  magnetisation along the $c$ axis in their vicinity and several of them have a net magnetic moment. Although the weak ferromagnetic moment -- arising from canting of Mn spins -- is small, it can be significantly enhanced by the magnetisation of rare-earth ions coupled to Mn spins, as in the bulk A$_2$ phase.\cite{Hisashi_JPSJ_2002}

In conclusion, multiferroic hexagonal manganites provide a rich playground for physics of topological defects in multiple coexisting orders. We presented a theory based on first-principles calculations that explains the observed coupling between the structural distortions, electric polarisation and spins at the domain walls and vortices. The significance of these findings lies in the fact that topological defects can dominate cross-coupling responses of bulk materials, such as magnetoelectric switching.

SA and MM were supported by the ZIAM Groningen under award MSC06-20 and by FOM grant 11PR2928. 
KTD acknowledges fellowship support from the International Center of Materials Research and computational resources
provided by the CNSI Computing Facility at UC Santa Barbara through Hewlett-Packard.
NS was supported by the ETH Z\"{u}rich.

\appendix
\section{First-principles calculations}

We performed \emph{ab initio} calculations using Kohn-Sham density functional theory (DFT), as implemented in the 
\verb!ABINIT!\footnote{The ABINIT code is a common project of the Universit\'e Catholique de Louvain, 
Corning Incorporated, and other contributors (URL http://www.abinit.org).} software 
package\cite{Abinit1,Abinit2}.
All calculations employ a supercell approach with periodic boundary conditions.
Wave functions and charge densities are expanded in a plane-wave basis. 
Efficient computational treatment of heavy elements is 
achieved using the projector-augmented wave method for core-valence partitioning\cite{AbinitPAW},
which significantly reduces the required plane-wave energy cutoff.

We approximate the Kohn-Sham exchange-correlation potential using the local spin density 
approximation\cite{LSDA} with a Hubbard-U correction applied to the partially-filled manganese $d$ states following the Liechtenstein
approach\cite{AbinitPAWU} with double-counting corrections in the fully localised limit. 
All calculations reported here were performed with values of $U=8.0$\,eV for the Coulomb integrals and $J=0.88$\,eV for the
intra-atomic exchange coupling, as chosen previously by Fennie and Rabe\cite{Fennie_PRB_2005}.
We enforce an A-type antiferromagnetic ordering for all calculations\cite{Fennie_PRB_2005}.
With this choice of parameters and magnetic ordering, an insulating electronic structure in the high-symmetry
(P6$_3$/mmc) crystal structure results, with a Kohn-Sham band gap of $0.75$\,eV.
Since this underestimates experimentally reported values of the band gap,\cite{Smith_JACS_2009} 
we carefully verified 
that no spurious metal-insulator transitions occur as structural distortions are introduced, so 
that the free-energy landscape contains no anomalous features.

The parameters given in Eq.~(\ref{eq:Fu}) can be extracted by considering homogeneously
distorted periodic structures that are commensurate with the wave vectors of all distortions. 
The smallest unit cell that can accommodate all possible values of $Q$, $\Phi$ and ${\cal P}$ contains $30$ atoms\footnote{We note that we
do not impose directly a constraint on the value of $P_c$. Rather, we impose the magnitude of the $\Gamma_2^-$ mode
on the structure, which results in $P_c \neq 0$.}, or $6$ formula units of YMnO$_3$.  
To fit the parameters, we explore the variation of the DFT total energy with the magnitude of $Q$, $\Phi$ and ${\cal P}$. 
The ionic positions are defined by the projection of the DFT ground-state P6$_3$cm structure onto the
$\Gamma_2^-$ and $K_3$ modes of P6$_3$/mmc combined with the desired value of  
$Q$, $\Phi$ and ${\cal P}$.
However, we fully optimise the cell parameters for each mode distortion to eliminate stresses. 
The result is that the homogeneous contribution to stresses have been implicitly eliminated through renormalisation of the 
Landau parameters.

The dependence of the free energy on ${\cal P}$ with $Q=0$ is shown in Fig.~\ref{fig:dft_homogfits}(b), reaffirming
the stability of the polar mode in the high-symmetry structure, and therefore the improper nature
of the ferroelectricity in YMnO$_3$. Fig.~\ref{fig:dft_homogfits}(a) and (c) respectively show computations for 
the trimerisation mode and the coupling between trimerisation and polar modes.

\begin{figure}[htbp]
\centering
\includegraphics[width=1.0\columnwidth]{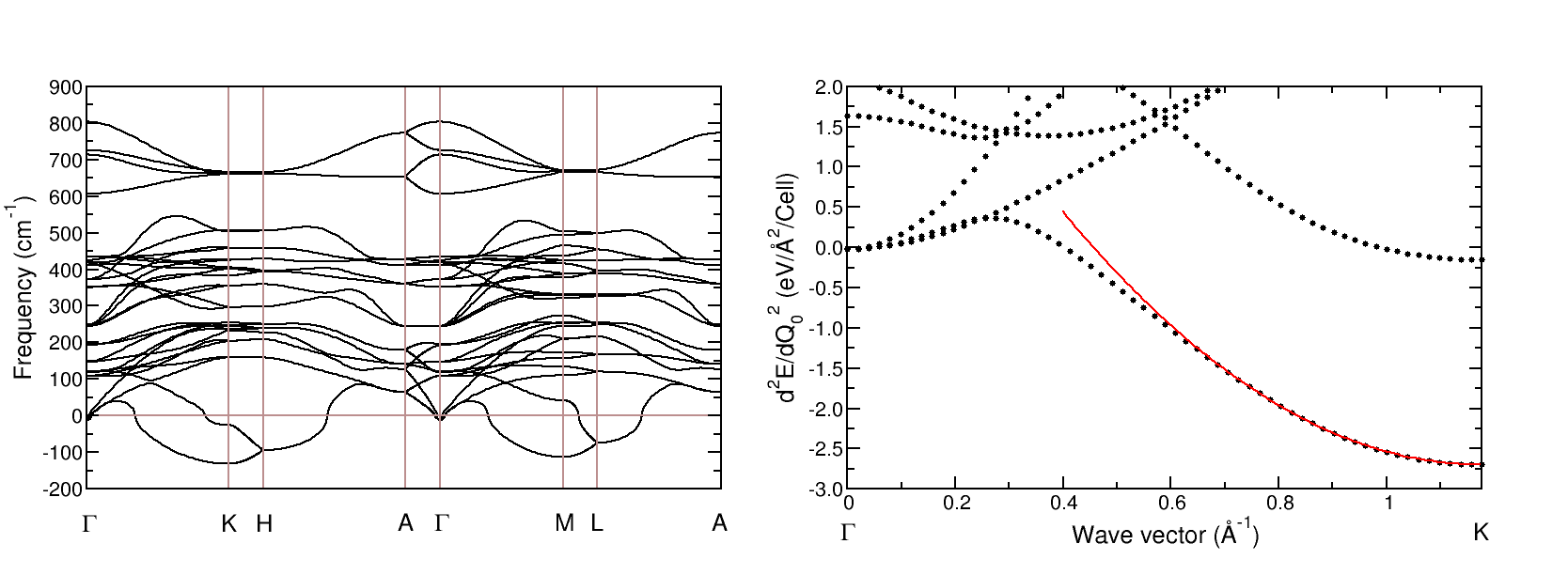}
\caption{Left panel: Phonon band structure of the high-symmetry phase (P6$_3$/mmc) of YMnO$_3$, obtained 
using frozen-phonon methods and Fourier interpolation of the interatomic force constants. The most unstable 
zone-boundary mode at $K$ is the $K_3$ trimerisation mode. 
Right panel: Extraction of the $s^x_Q$ stiffness parameter by fitting the dispersion of the 
force constant of the trimerisation mode. }
\label{fig:phonstiff}
\end{figure}

With all homogeneous Landau parameters given in Eq.~(\ref{eq:Fu}) specified, we now turn our attention to
the stiffness parameters. 
The most convenient way to proceed is to write the spatial inhomogeneity in $Q$ and ${\cal P}$ as a single harmonic, 
for example $Q\left(\vec{r}\right) = Q_q e^{i\vec{q}.\vec{r}}$ with $\Phi=0$. 
The stiffness energy has then the form
\begin{equation}
f_s\left(\vec{q}\right) = \frac{1}{2} \sum_i q_i^2\left[ s^i_Q \vert Q_q\vert^2 + s^i_P \vert{\cal P}_{q}\vert^2\right].
\end{equation}
This expression is harmonic in mode amplitudes $Q_q$ and ${\cal P}_{q}$.
Note that in order to compute $s^i_Q$, we do not need to consider spatial variations in the $\Phi$ field.
The computational cost of explicitly computing $f_s$ by imposing various short wave vectors, $q$, in our supercell density-functional calculations would be prohibitive. 
Instead we extract these harmonic terms using the method of frozen phonons combined with
Fourier interpolation of the inter-atomic force constants, a method commonly used for computing phonon 
band structures. 
Fig. \ref{fig:phonstiff} shows our calculated phonon band structure for the high-symmetry structure (P6$_3$/mmc) of YMnO$_3$. 
The strongest instability at the $K$ point is the cell-tripling $K_3$ trimerisation mode. 
As previously noted, all $\Gamma$ phonons, including the $\Gamma_2^-$ polar mode, are stable in the high-symmetry structure.

Multiplying the dynamical matrix used in the computation of phonon modes by the weighted mass, we obtain a 
$\vec{q}$-dependent force-constant matrix:
\begin{equation}
C_{i,j}\left(\vec{q}\right) = \sqrt{M_i,M_j} D_{i,j}\left(\vec{q}\right),
\end{equation}
and by identifying the relevant branch, the $\vec{q}$-dependent 
eigenvalues of $C_{i,j}\left(\vec{q}\right)$ are related to 
$\frac{\partial^2 f_s}{\partial Q_0^2}$ or $\frac{\partial^2 f_s}{\partial {\cal P}_{0}^2}$. 
Hence, the required stiffness parameters can be extracted. 
As an example, Fig. \ref{fig:phonstiff} shows the extraction of $s^x_Q$ by fitting the 
$q_x$ dispersion of the force constant of the unstable trimerisation branch. 
Using this technique, we find the stiffness parameters for YMnO$_3$ listed in Table~\ref{tab:parameters}.

\end{document}